\documentclass[prl,aps,superscriptaddress,twocolumn
,floatfix]{revtex4}
\usepackage{graphicx}
\usepackage{amssymb}
\usepackage[hypertex]{hyperref}

\begin{document}
\advance\textfloatsep by -0.15in
\advance\abovecaptionskip by -0.15in
\def\topfraction{1.00}
\def\textfraction{0.0}

\title{Supersolids versus phase separation in two-dimensional lattice bosons}

\author{Pinaki Sengupta}
\affiliation{Department of Physics, University of California, Riverside, CA
  92521}
\author{Leonid P. Pryadko}
\affiliation{Department of Physics, University of California, Riverside, CA
  92521}
\author{Fabien Alet}
\affiliation{Theoretische Physik, ETH Z\"urich, CH-8093 Z\"urich, Switzerland}
\affiliation{Service de Physique Th{\'e}orique, CEA Saclay, F-91191
  Gif sur Yvette, France} 
\author{Matthias Troyer}
\author{Guido Schmid}
\affiliation{Theoretische Physik, ETH Z\"urich, CH-8093 Z\"urich, Switzerland}
                                                                              
\date{\today}

\begin{abstract}
  We study the nature of the ground state of the strongly-coupled two
  dimensional extended boson Hubbard model on a square lattice.  We
  demonstrate that strong but finite on-site interaction $U$ along
  with a comparable nearest-neighbor repulsion $V$ result in a
  thermodynamically stable supersolid ground state just above
  half-filling, and that the checker-board crystal is unstable for smaller
  $V$, and for any $V$ just below half-filling. The interplay
  between these two interaction energies re\-sults in a rich phase
  diagram which is studied in detail using quantum Monte Carlo methods.
\end{abstract}

\pacs{75.40.Gb, 75.40.Mg, 75.10.Jm, 75.30.Ds}

\maketitle

The detection of possible supersolid (SS) state in recent experiments
on solid $^4$He by Kim and Chan \cite{kc-nat} has led to
renewed interest \cite{leggett-2004} in a
problem that has long \cite{penrose} intrigued physicists: Can a
supersolid phase---with simultaneous diagonal (solid) and off-diagonal
(superfluid) long-range order---exist in a bosonic system?  
While the the issue remains
controversial \cite{andreev-chester-leggett,anderson,meisel} in a
translationally invariant system despite almost fifty years of
theoretical research,  
the situation in lattice models is clearer.

Theoretical studies \cite{mft1,balents,otterlo} of various {\em
  lattice\/} boson models (which can nowadays be implemented using
cold bosonic atoms on optical lattices \cite{optical-latt}), appeared
to confirm that here supersolid ground states can indeed exist,
particularly when doped away from half-filling.  Studies of the
closely related quantum phase model found supersolid order in the
ground state even at half-filling \cite{otterlo}.  However, as was pointed out
recently the stability of the supersolid against phase
  separation had not been investigated \cite{rts-ps}.  Indeed, for
hard-core bosons on a square lattice, the most widely discussed
supersolid pattern---with $(\pi,\pi)$ diagonal order---is
thermodynamically unstable and phase separates into a pure $(\pi,\pi)$
solid and a superfluid (SF) for all values of interaction strengths.
A striped supersolid phase---with $(0,\pi)$ ordering---is stabilized
by a finite next-nearest-neighbor (nnn) interaction.

\begin{figure}[tb]
  \includegraphics[width=7.5cm]{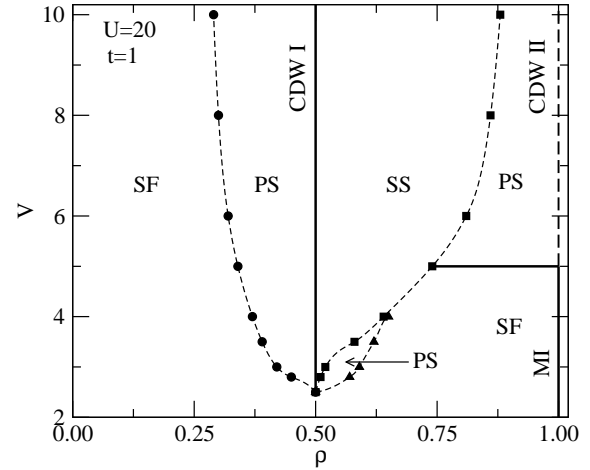}
  \caption{The ground state phase diagram of the 2D extended Bose-Hubbard
    model (\protect{\ref{eq:H}}) in the $V-\rho$ plane for $U/t=20$
    and densities $\rho\le1$, showing superfluid (SF) phases,
    checkerboard solids formed by single bosons (CDW I) and pairs of
    bosons (CDW II), a Mott-insulating phase (MI), phase separation
    (PS) and finally a supersolid phase (SS).} 
\label{fig:vn}
\end{figure}

In this work we analyze stability of crystalline and supersolid orders
of lattice bosons.  We present exact strong-coupling arguments showing
under which conditions checkerboard supersolids are unstable, and how
they can be stabilized with large but finite on-site and
nearest-neighbor (nn) energies $U$ and $V$.  We support
these arguments by quantum Monte Carlo simulations of a
two-dimensional (2D) extended Bose Hubbard model demonstrating that
the supersolid (SS) phase is stabilized for densities $\rho>1/2$ and
sufficiently large $V$ (Fig.~\ref{fig:vn}).

Specifically, we study the extended Bose-Hubbard model (EBHM) on
a $d$-dimensional hypercubic lattice  with
on-site ($U$) and nn ($V$) interactions, 
\begin{eqnarray}
H &=&-t \sum\nolimits_{\langle i,j\rangle}
  (a_i^{\dagger}a_j+a_j^{\dagger}a_i)-\mu\sum\nolimits_i n_i \nonumber\\
& & +V\sum\nolimits_{\langle i,j\rangle} n_in_j+{U\over 2}
\sum\nolimits_i n_i\,(n_i-1), 
\label{eq:H}
\end{eqnarray}
where $a_i^{\dagger}(a_i)$ creates (annihilates) a boson at site $i$
with the occupation number $n_i\equiv a_i^\dagger a_i$, $t$ is
hopping, $\mu$ is the chemical potential, and $\langle i,j\rangle$
runs over all nn pairs. 

In the zero-hopping limit, $t=0$, the non-negative potential energy
($U, V>0$) is minimized at
half-filling, $\rho=1/2$, by the crystal state with only one
sublattice occupied [checkerboard pattern with $(\pi,\pi)$ modulation
in 2D].  This state is gapped; it remains stable in the presence of a
small hopping, $t\ll U,V$, with a kinetic energy gain $\Delta E\approx
-zt^2/[(z-1)V]$ per boson, where $z=2d$ is the coordination number.

\begin{figure}[t]
  \includegraphics[width=2.3in]{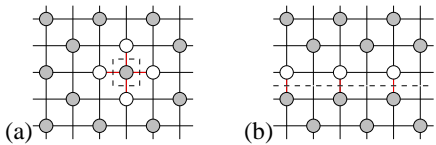}
  \caption{The $\rho=1/2$ checkerboard crystal doped with holes.  (a)
    Four holes encircle a boson which can hop between the five
    degenerate sites.
    (b)  Domain wall doped with holes; bosons
    can hop freely across the dashed line.} 
\label{fig:holes}
\end{figure}

Introducing holes  only costs chemical potential $\mu$ but no
potential energy; the kinetic 
energy gain is somewhat increased but remains quadratic in $t$ for
isolated holes.  However, the kinetic energy gain becomes {\em
  linear\/} in $t$ if a number of holes encircle a region of a crystal
[Fig.~\ref{fig:holes}(a)].  The energy gain is maximized at $\Delta
E\approx -ct$, $1<c<2$ per hole for a planar [linear in 2D, see
  Fig.~\ref{fig:holes}(b)] domain wall doped with one hole per two
sites.  As a result, for a large system with $N=L^d$ sites, the
crystalline order is destroyed by introduction of a small density
$\rho\sim L^{-1}$ of holes.  This instability of the $\rho=1/2$
crystal to domain wall formation upon hole doping excludes the
possibility of a SS phase.  In practice, on the isotropic square
lattice, the instability develops further, leading to a phase
separation between the commensurate crystal at $\rho=1/2$ and a
uniform superfluid with $\rho<1/2$ (Fig.~\ref{fig:vn}).

Doping of the $\rho=1/2$ crystal with additional bosons works 
differently depending on the relation between $V$ and $U$.  The energy
cost to place a boson at an empty (occupied) site is $E_0\equiv
zV-\mu$ ($E_1\equiv U-\mu$).
Respectively, for $U>zV$, the additional bosons fill empty sites and
mask the checkerboard modulation; for $U-zV\gg t>0$ the
situation is precisely particle-hole conjugate to hole doping.  The
kinetic energy is again minimized at planar domain walls which
destabilize the checkerboard crystal order.  In particular, in the
hard-core limit $U\to\infty$, the crystalline order is always unstable
for $\rho\neq1/2$.

With $zV\gtrsim U$, however, the bosons (unlike holes) can be placed
on either an occupied or unoccupied site.  The total energy of a
single boson delocalized between the two sublattices is
$E=E_1+\Delta-(2z^2t^2+\Delta^2)^{1/2}$, where $\Delta\equiv (zV-U)/2$.
Clearly, for sufficiently small $\Delta\sim t$, the kinetic energy
$E-E_1$ is again linear in $t$ and large, which prevents the domain wall
formation.  As a result, these doped particles will form a superfluid
on top of the charge ordered background and hence a supersolid.  Two
bosons experience both on-site and nn repulsion ($2U$ and $V$
respectively).  Therefore, at sufficiently small densities
the condensate should remain stable, which completes the formal
argument for the supersolid existence.

Similarly, at unit filling, $\rho=1$, the ground state is a Mott
insulator with one boson per site for $U>zV$, and an ordered solid
with two bosons on every other site for $U<zV$.  In the former case,
additional holes (particles) move along the uniform background with
the hopping integral $t$ ($2t$) and experience both nn and on-site
repulsion (infinite in the case of holes).  They condense on top of
the uniform background forming a superfluid.  However, for $zV>U$, the
doped particles move on a checkerboard background with the effective
hopping, e.g., $t_*=2t^2/[zV-U+(z-2)V]$ for holes.  The resulting
kinetic energy gain is only quadratic in $t$ and can be superseded if
the holes come together into a supersolid phase with $\rho\gtrsim1/2$.
Overall, this leads to a thermodynamical instability of the hole-doped
checkerboard solid formed by pairs of bosons at $\rho=1$: the system
can minimize its energy by phase separation.  Note that here phase
separation is not between a superfluid and a solid but between a
supersolid and a solid. The solid order is not destabilized at this
first order phase transition, but just the ``Bose-Einstein
condensation'' transition of holes doped into the solid becomes first
order. 

We next perform quantum Monte Carlo simulations to corroborate these
arguments and to show the phase diagram and the existence of a
supersolid phase for the EBHM in the low-density region $\rho\le1$.

We have used loop-operator updates in a stochastic series expansion
(SSE) quantum Monte Carlo (QMC) method \cite{sse2} to study the EBHM
(\ref{eq:H}) in the strong coupling regime ($U,V \gg t$) and for
$0<\rho \le 1$. In the present study, simulations have been carried
out in square geometry, 
$N=L\times L$, with $L=6,\ldots, 16$.  Ground state properties have
been obtained by taking 
sufficiently large values of the inverse temperature $\beta$, where
$\beta=2L$ turned out to be sufficient.

To characterize different phases, we have studied the static
staggered [${\bf Q}=(\pi,\pi)$] structure factor,
\begin{equation}
  S({\bf Q})={1\over N}\sum_{j,k}e^{-i{\bf Q}\cdot({\bf
           r}_j-{\bf r}_k)}\\ 
       \langle  n_jn_k\rangle - \langle n_j\rangle^2, 
\end{equation}
which measures the diagonal long range order (checkerboard solid) in
the system,  and 
the superfluid density $\rho_s$, measured from the winding numbers of the
bosonic world lines ($W_x$ and $W_y$) in the $x$- and $y$- directions 
as $\rho_s=\langle W_x^2+W_y^2\rangle/2\beta m$, where $m=2/t$ is the 
effective mass of the bosons.  A checkerboard solid ground state at
$\rho=0.5$ is marked by a diverging $S(\pi,\pi)$ and vanishing
$\rho_s$, whereas a pure superfluid phase has $S(\pi,\pi)=0$ and
$\rho_s>0$. A supersolid phase, on the other hand, is characterized by
a diverging $S(\pi,\pi)$ {\em and\/} a non-zero value of $\rho_s$. For finite
system sizes, both quantities are always finite and estimates for the
thermodynamic limit are obtained by carefully studying the finite-size
scaling of the observables. 

A jump in $\rho$ with varying $\mu$ indicates a
discontinuous (first order) transition, and has been used to identify
regions of phase separation in the canonical ensemble (fixed density
$\rho$). We postpone a more rigorous analysis to accurately identify
the nature of the transitions and the accurate domain boundaries to a
later study, and focus, instead, on establishing  the existence of SS
phase over finite regions of the parameter space. 

\begin{figure}[thb]
  \includegraphics[width=7.0cm]{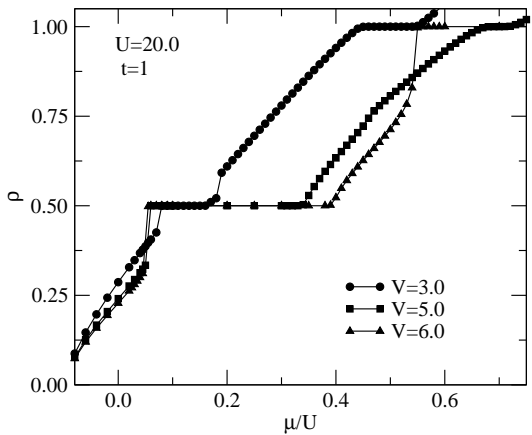}
  \caption{The average density as a function of the chemical potential
    for three different values of $V$ ($U=20$). For clarity of
    presentation, data for only one system size, $L=16$, is shown.
    Error bars are smaller than the symbol sizes. Discontinuous
    transitions are marked by finite jumps in the particle density.}
  \label{fig:avn}
\end{figure}

\begin{figure}[thb]
  \includegraphics[width=3.09in]{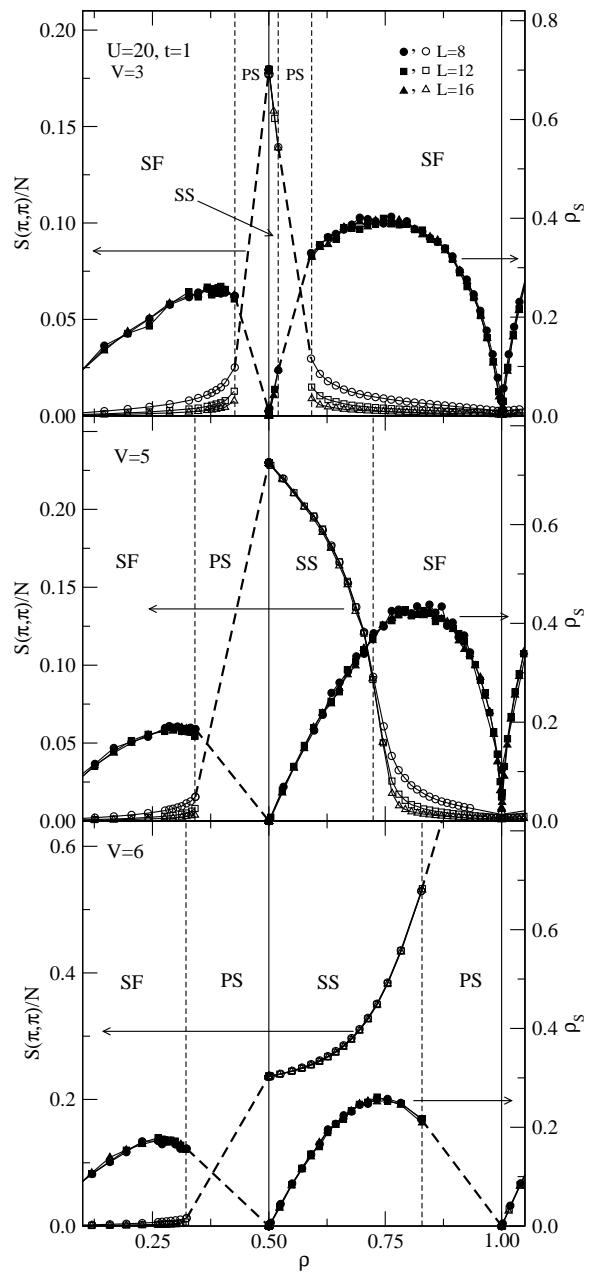}
  \caption{The scaled static staggered structure factor (open symbols) and 
    superfluid stiffness (filled symbols) as a function of average
    particle density $\rho$ for three representative values of $V$. At
    small $\rho$, the ground state is a superfluid (SF) for all $V$.
    With increasing density there is discontinuous transition to a
    ($\pi,\pi$)-ordered charge density wave (CDW) at $\rho=0.5$, with
    an intermediate region of phase separation (PS). At $\rho \gtrsim
    0.5$, the ground state is a supersolid (SS) with finite values of
    both $S(\pi,\pi)$ and $\rho_s$.  For $V<U/4$, this is followed by
    another discontinuous transition to a SF (the precise mapping of
    the SS/PS boundary requires larger $L$), with a second region of
    PS. For $V=U/4$, the SS region extends to
    higher densities with a continuous transition to a SF ground state
    at $\rho\approx 0.74$.  The SS state extends to even higher
    densities for $V>U/4$, but is followed by a discontinuous
    transition to a ($\pi,\pi$)-ordered CDW at $\rho=1$, with an
    accompanying region of PS. For $V\le U/4$, the ground state at
    $\rho=1$ is a Mott insulator.}
\label{fig:combo}
\end{figure}

A plot of $\rho$ as a function of $\mu$ shows clear indications of 
phase separation at $\rho< 0.5$ for all values of $V$---the
discontinuity in $\rho$ grows with increasing $V$. For $\rho > 0.5 $,
the curves are qualitatively different. For $V<U/4$, there is a small,
but finite, region of positive slope (eg., $0.5 < \rho < 0.52$ for
$V=3$), 
followed by phase separation for $0.52 < \rho < 0.60$, and a region 
of positive
slope for $\rho > 0.6$. At $V=U/4$, there is no evidence of phase
separation for $\rho > 0.5$. With $V>U/4$, the region of phase
separation shifts to large densities, $\rho \lesssim 1$.  The location
and extent of phase separated regions for small $V (<U/4)$ agrees well
with the results of Ref.~\onlinecite{rts-ps}b apart from the extra
region of positive slope 
for $0.5< \rho < 0.52$. As shown below, the ground state at
these densities has supersolid order.  The extent of the supersolid
phase decreases rapidly with decreasing $V$, becoming vanishingly
small in the limit $V\ll U $.  We note that for small $V$ the excess
density $\rho-1/2<1/L$ and larger $L$ are required to 
map the SS phase boundary accurately.  

Ground state results for $S(\pi,\pi)$ and $\rho_s$ as a function of
$\rho$ for three representative values of $V$ are shown in
Fig.~\ref{fig:combo} for three different system sizes $L$. The data 
are seen to be well converged with system size.  At
small $\rho$, the ground state is a superfluid (SF)---the stiffness 
converges to a finite value while $S(\pi,\pi)$ scales to zero.
As the density increases beyond a critical value $n_{c1}$, there
is a discontinuous transition to a ($\pi,\pi$) ordered
charge-density-wave (CDW) ground state with $\rho=0.5$. Any
intermediate density is inaccessible in the grand canonical ensemble.
For $V=3t(<U/4$), at $\rho>0.5$, there are indications for a small
region of supersolid (SS) 
characterized by finite values of {\em both\/} $S(\pi,\pi)$ and
$\rho_s$,  but further finite size scaling tests will be needed to
check whether this region remains in the thermodynamic limit. With
increasing density, there is another discontinuous 
transition to an SF state with a second region of PS. Finally, at
$\rho=1$, the ground state is a Mott insulator (MI) with both
$S(\pi,\pi)=0$ and $\rho_s=0$. For $V=5t(=U/4)$, the extent of the SS
region increases substantially and its stability is well
established. Additionally, the second phase 
separated region shrinks to zero and there is a direct SS-SF
transition. For $V=6t (>U/4)$, the SF phase at high densities is
replaced by another region of phase separation. The ground state at
$\rho=1$ changes from a MI to a ($\pi,\pi$) ordered CDW state with two
particles occupying every alternate lattice site, with a discontinuous
transition separating it from the SF state.

\begin{figure}
  \includegraphics[width=7.5cm]{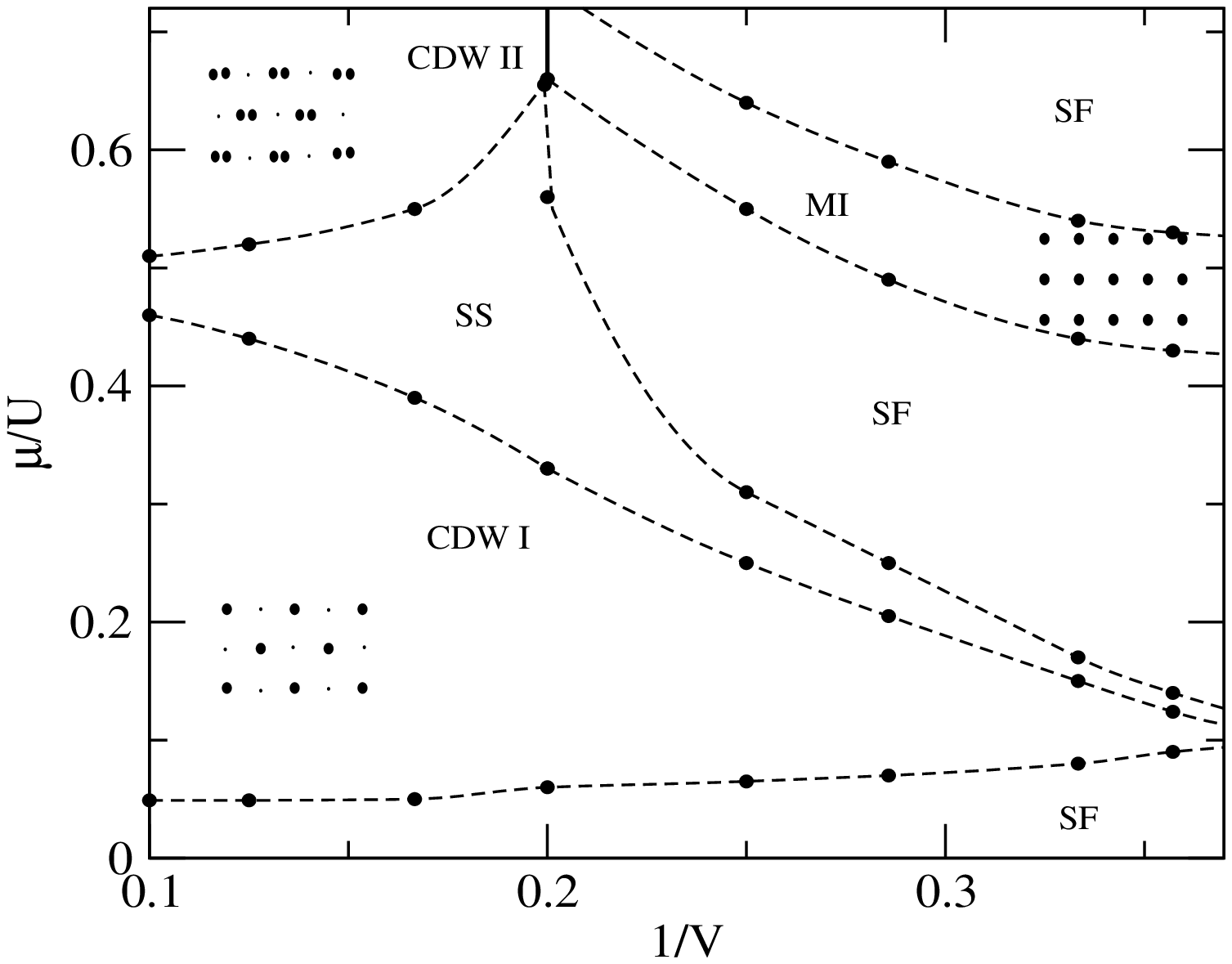}
\caption{The ground state phase diagram in the $V-\mu$ plane,
  notations as in Fig.~\protect\ref{fig:vn}.  Different 
  solid-ordered phases are shown schematically. The PS regions are
  manifested as discontinuous transitions across the corresponding
  phase boundaries (not shown).}
\label{fig:vmu}
\end{figure}

The results are combined to map the schematic ground state phase
diagram of the Hamiltonian (\ref{eq:H}) in the $V-\rho$
(Fig.~\ref{fig:vn}) and $\mu-1/V$ (Fig.~\ref{fig:vmu}) planes.
Fig.~\ref{fig:vn} shows the different phases in the ($V,\rho$)
parameter space at a constant value of the on-site interaction,
$U=20$, $t=1$.  For small $V$, the ground state is a SF for all $\rho
< 1$. At $V > 2.5t$, the different phases appear as shown in the
figure. The extent of the supersolid phase and that of the phase
separated region at $\rho < 0.5$ increases with increasing $V$,
whereas the phase separated region at $\rho > 0.5$ gets vanishingly
small for moderate values of $V$. It is not clear from the available
data if the PS--SF phase boundary meets the SS boundary at a point, or
approaches it asymptotically. At $V>5.0t$, the SF region at high
densities is replaced by a phase separated region, while the ground
state at $\rho=1$ changes from a MI to a ($\pi,\pi$)-ordered CDW with
two bosons occupying every other lattice site. It should be emphasized
that the phase diagram is qualitative and the phase boundaries are
approximate.

The features of the phase diagram as a function of $1/V$
(Fig.~\ref{fig:vmu}) are markedly different from the ``lobe''
structure observed in a plot of $\mu$ as a function of $t/U$
for the EBHM. The nature of the
ground state at $\rho=1$ changes from a CDW to an MI as $V$ is varied
across $U/4$. This is accompanied by a change in the curvatures of the
phase boundaries. Furthermore, the MI region remains finite even in
the limit of $V\rightarrow 0$. No evidence of supersolid phase is
found at $\rho=0.5$, in agreement with the variational studies and
previous numerics \cite{mft1,otterlo}.

In conclusion, we have used exact strong-coupling expansion and QMC
simulations to study the nature of the ground state phases of the
extended boson-Hubbard model on a square lattice. The interplay of the
on-site and nearest-neighbor interactions leads to a rich phase diagram
including a supersolid phase with simultaneous diagonal and
off-diagonal long-range order.  We have provided strict arguments why
a soft-core model with $V>U/z$ and densities $\rho>1/2$ is sufficient
to stabilize a supersolid phase in a model with nearest neighbor
couplings only.  This is in contrast to the ``hard-core'' bosons where
the system phase separates for all values of the nn interaction
strength and additional nnn interactions  or hoppings are
  needed to stabilize a supersolid ground state.  Also, in the
studied range of parameters (including very large $V<12t)$ we have not
found any nominally gapless phase with both $S(\pi,\pi)$ and $\rho_s$
zero which could 
potentially be identified with a bose metal \cite{bose-metal} (see
also the argument against such a phase in  Ref.~\onlinecite{rts-ps}d).

It is a pleasure to thank N.\ Prokof'ev, B.\ Svistunov, and C.\ M.\ 
Varma for useful discussions and R.\ T.\ Scalettar for suggesting the
problem. Simulations were carried out on the computer cluster at the
Institute of Geophysics and Planetary Physics at the University of
California, Riverside.


\begin{thebibliography}{19}
\bibitem{kc-nat}
E. Kim and M. H. W. Chan, Nature {\bf 427}, 225 (2004); Science {\bf
  305}, 1941 (2004). 

\bibitem{leggett-2004}
A. Leggett, Science {\bf 305}, 1921 (2004); N. Prokof'ev, and
B. Svistunov, preprint cond-mat/0409472; 
A. S. Moskovin, I. G. Bostrem, and A. S. Ovchinikov,  preprint
cond-mat/0404561. 

\bibitem{penrose}
O. Penrose, and L. Onsager, Phys. Rev. {\bf 104}, 576 (1956).

\bibitem{andreev-chester-leggett}
A. F. Andreev, and I. M. Lifshitz, Sov. Phys. JETP {\bf 29}, 1107 (1960);
G. Chester, Phys. Rev. A {\bf 2}, 256 (1970);
A. J. Leggett, Phys. Rev. Lett. {\bf 25}, 1543 (1970).

\bibitem{anderson}
P. W. Anderson, {\em Basic notions of Condensed Matter Physics\/}
(Benjamin, New York, 1984).

\bibitem{meisel}
For review of experimental and theoretical work see M. W. Meisel, Physica 
{\bf 178B}, 121 (1992), and other articles in the same volume.

\bibitem{mft1}
H. Matsuda, and T. Tsuneto, Suppl. Prog. Theor. Phys. {\bf 46}, 411 (1970);
K. S. Liu, and M. E. Fisher, J. Low Temp. Phys. {\bf 10}, 655 (1973);
E. Roddick and D. Stroud, Phys. Rev. B {\bf 48}, 16600 (1993); 
{\em ibid.}\ {\bf 51}, 8672 (1995). G. G. Batrouni {\it et al.},
Phys. Rev. Lett. {\bf 74}, 2527 (1995); R. T. Scalettar {\it et al.},
Phys.  Rev. B {\bf 51}, 8467 (1995); R. Mincas, S. Robaszkiewicz, and
T. Kostyrko, {\em ibid.}\ {\bf 52}, 6863 (1995); E. S. Sorensen and
E. Roddick, {\em ibid.}\ {\bf 53}, R8867 (1996).

\bibitem{balents}
  E. Frey and L. Balents, Phys. Rev. B {\bf 55}, 1050 (1997).

\bibitem{otterlo}
A. van Otterlo and K.-H. Wagenblast, Phys. Rev. Lett. {\bf 72}, 3598 (1994);
A. van Otterlo {\em et al.}, Phys. Rev. B {\bf 52}, 16176 (1995).

\bibitem{optical-latt} 
M. Greiner {\em et al.}, Nature {\bf 415}, 39 (2002).

\bibitem{rts-ps}
Kohno and Takahashi,  Phys. Rev. B {\bf 56}, 3212 (1997);
G. G. Batrouni, and R. T. Scalettar, Phys. Rev. Lett. {\bf 84}, 1599 (2000);
H. H{\'e}bert {\em et al}., Phys. Rev. B {\bf 65}, 014513 (2001);
A. Kuklov, N. Prokof'ev, and
B. Svistunov, Phys. Rev. Lett. {\bf 93}, 230402 (2004).


\bibitem{sse2}
A. W. Sandvik, Phys. Rev. B {\bf 59}, R14157 (1999). 

\bibitem {bose-metal} 
P. Phillips, and D. Dalidovich, Science {\bf 302}, 243 (2003); 
S. Das, and S. Doniach, Phys. Rev. B {\bf 64}, 134511 (2001); 
{\em ibid.}\ {\bf 60}, 1261 (1998).


\end{thebibliography}
\end{document}